\documentclass[twocolumn,onecolappendix,tighten]{aastex701}
\usepackage{amssymb,amsmath,framed}

\usepackage{bm}
\expandafter\ifx\csname package@font\endcsname\relax\else
 \expandafter\expandafter
 \expandafter\usepackage
 \expandafter\expandafter
 \expandafter{\csname package@font\endcsname}
\fi
\begin{document}

\title{Undetected Past Contacts with Technological Species: Implications for Technosignature Science}

\author[orcid=0000-0001-7807-7073, sname='Grimaldi']{Claudio Grimaldi} 

\affiliation{Laboratory of Statistical Biophysics, Ecole Polytechnique F\'ed\'erale de Lausanne -- EPFL,1015 Lausanne, Switzerland}
\affiliation{Centro Studi e Ricerche Enrico Fermi -- CREF, Piazza del Viminale, 1, 00184 Roma, Italy}
\email[show]{claudio.grimaldi@epfl.ch}

%\keywords{\uat{Search for extraterrestrial intelligence}{2127} --- \uat{Technosignatures}{2128} --- \uat{Astrobiology}{74}}

\begin{abstract}
In the search for extraterrestrial intelligence (SETI), the highly incomplete sampling of the technosignature search space is often considered 
as a plausible explanation for the persistent lack of detections over six decades of searches. If correct,
this would imply that technosignatures may already have reached Earth without being detected or correctly identified.
Here, we explore this possibility using a Bayesian inference framework to estimate present-day detectability given $n\ge 1$
undetected contacts over the past $65$ years -- the period since the first SETI experiment. 
We show that achieving high detectability of technosignatures emitted within a few hundred light-years of Earth would require implausibly
large $n$ values, even exceeding the population of habitable planets within that range. More conservative estimates 
can be obtained only assuming that emitters are tightly clustered near Earth or that their population in the Milky Way 
has undergone a very recent and sudden boost. This tension is further exacerbated for short-lived technosignatures and persists
whether they are omnidirectional, as in Dysonian megastructures, or directional, as in intentional communication attempts. 
These findings suggest that, if undetected past contacts from the Milky Way have indeed occurred, the best prospects of detection may 
lie in searches extending over several thousand light-years, though only a few detectable technoemissions would be expected. 
\end{abstract}

\section{Introduction}
\label{SecIntro}

The search for technosignatures -- remotely detectable indicators of extraterrestrial technology -- has been ongoing for 
over sixty years, beginning with the first SETI experiment in 1960 \citep{Drake1961}. Since then, more than 200 searches have been 
conducted\footnote{\url{https://technosearch.seti.org/}}, targeting predominantly
artificial radio signals, as well as optical and infrared emissions from beyond the Solar System, albeit less frequently
and for only about fifty years. To date, however, 
no confirmed detections have been reported.

This lack of detection does not necessarily imply that humanity is the only technological civilization in the Galaxy.
Rather, it may reflect that only a minute fraction of the vast multidimensional parameter search space has been explored
so far \citep{Tarter2010,Wright2018} -- a situation compared to searching for a needle in a cosmic haystack \citep{Wright2018}. 
Consequently, the question of the prevalence, or even the existence, of advanced technological species in our Galaxy 
remains unresolved.

Even so, the absence of detections constitutes a valuable data point. While it does not constrain the existence
of extraterrestrial technologies \textit{per se}, it does inform us about upper limits on the prevalence of their detectable 
manifestations \citep{Enriquez2017,Price2020,Wlodarczyk2020,Gajjar2022,Marcy2022,Suazo2022,Margot2023,Garrett2023,Uno2023,Tremblay2024,Manunza2025,Mason2025}. 
In fact, to be potentially detectable, any technosignature must be observable from our vantage point at the time
of observation. In particular, electromagnetic technosignatures (hereafter, technoemissions) emitted at some point in the past from 
elsewhere in the Galaxy must intersect Earth -- an event referred to as ``contact" in what follows -- during the observational time window.  
Such a contact represents a necessary condition for potential 
detection \citep{Grimaldi2017,Balbi2018}. However, it is not a sufficient one, as factors such as
signal strength, frequency characteristics, and distinguishability from background noise may still prevent actual detection.

From this perspective, possible explanations for the decades-long absence of detections can be classified into two broad categories : either technoemissions are sufficiently rare that none has illuminated Earth during the relatively brief 65 years of SETI searches, 
or they have indeed intersected our planet but have so far eluded detection -- either because,
for instance, they occurred at unmonitored wavelengths, were below instrumental sensitivity thresholds, or were recorded but 
not recognized as artificial.

The first scenario posits that Earth has not been traversed by extraterrestrial technoemissions over six decades, 
thereby rendering their detection impossible. Under fairly general assumptions, this premise would imply fewer than one to five such emissions 
per century generated across the entire Galaxy \citep{Grimaldi2023}, suggesting that Earth is likely to remain unilluminated 
for at least several more decades. 

The second possibility depicts a more optimistic scenario. If undetected or unrecognized contacts with technoemissions have occurred in the 
recent past, it is reasonable to expect that others may still be ongoing. This would imply that the necessary condition for detection is 
more likely to be met, thereby increasing the chances of discovery as future searches become more comprehensive and effective.
Should the current, unprecedented commitment to the search for life elsewhere in general -- and for technosignatures in particular -- 
persists, this scenario suggests that a long-awaited discovery could be closer than ever.

Here, we quantitatively explore the implications of assuming that technoemissions generated across the Galaxy have intersected Earth since
the first SETI experiment $65$ years ago \citep{Drake1961}. Using a Bayesian framework, we show that even a few technoemissions 
detectable today within relatively short distances from Earth ($\lesssim 1,000$ ly) would imply 
unrealistically large numbers of undetected past contacts. 
Extending the analysis to the entire Milky Way reduces the number of past contacts required to sustain significant present-day detectability.
However, even in this case, detectable technoemissions are expected to be rare -- only a few across the entire Galaxy --
casting doubt on the prospect of a near-term detection.

By considering the possibility of past undetected contacts, this paper complements an earlier study \citep{Grimaldi2023}, in which the 
absence of confirmed detections to date was attributed to a complete lack of contacts with technoemissions throughout the history of SETI.

 \begin{figure*}[t]
	\begin{center}
		\includegraphics[width=0.8\textwidth,clip=true]{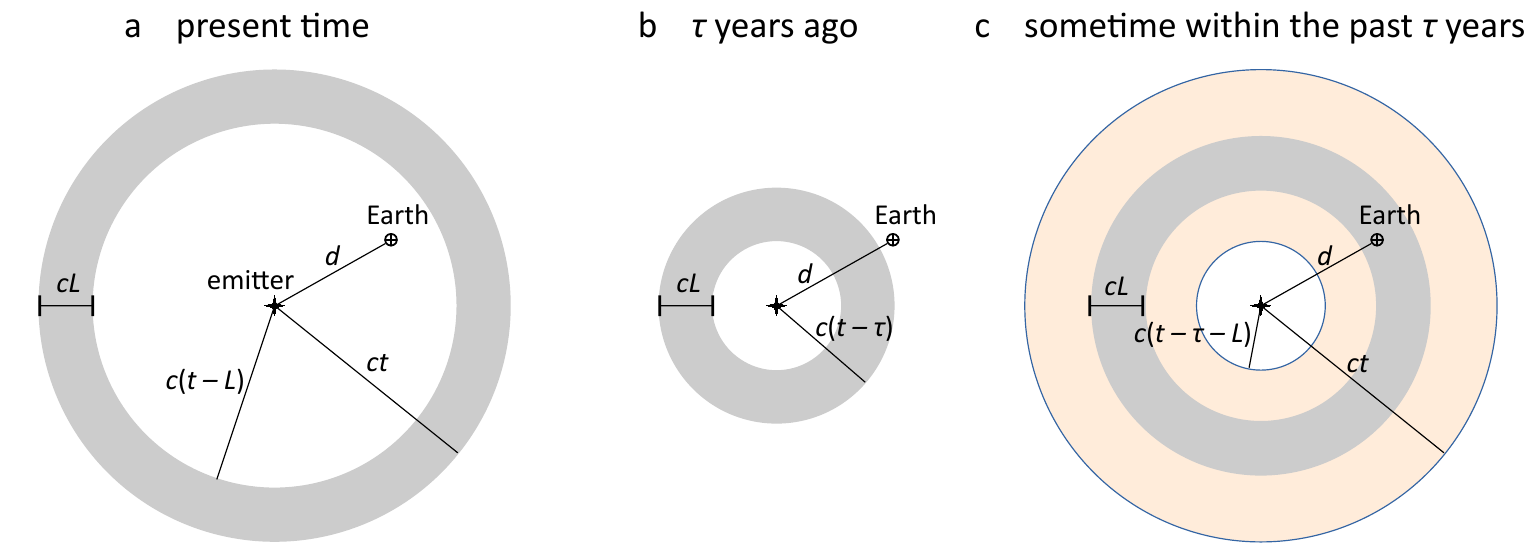}
		\caption{Temporal evolution of a spherical shell produced by an isotropic technoemission of duration $L$. In the present-day 
		configuration (panel a), the outer and inner radii of the shell are $ct$ and $c(t - L)$, respectively; the example shown corresponds 
		to the case in which Earth lies inside the hollow region of the shell, at a distance $d$ from the emitter. In panel b, the same shell is
		 shown $\tau$ years ago, when the outer and inner radii were smaller by $c\tau$ and Earth was outside the shell. In panel c, 
		 the shell has intersected Earth at some point over the past $\tau$ years if $c(t - \tau - L) \le d \le ct$. Expressed in terms of 
		 the emitter appearance time $t$, this condition becomes $d/c \le t \le d/c + L + \tau$, yielding an appearance time window for
		  intersection of $L + \tau$.
		}\label{fig1}
	\end{center}
\end{figure*}

\section{Model of technoemissions}
\label{model}
The model builds upon previous formulations \citep{Grimaldi2017,Grimaldi2018,Grimaldi2023}, which are briefly summarized below. We consider technoemissions to be generated within the Milky Way by technological species or their artifacts, hereafter
collectively referred to as ``emitters". The spatial positions of these emitters are specified by the random variable $\mathbf{r}$, denoting 
their vector position relative to the Galactic center. It is further assumed that emitters are statistically independent, with an appearance rate 
per unit volume, $\gamma(\mathbf{r},t)$, defined such that $\gamma(\mathbf{r},t)d\mathbf{r}dt$ corresponds to the expected number of emitters 
within the elemental volume $d\mathbf{r}$ centered at $\mathbf{r}$ that began emitting within a time interval $dt$ centered at time 
$t$ in the past (with $t>0$ measuring backward times). According to this definition, the quantity $\varGamma(t)=\int \! d\mathbf{r}\,\gamma(\mathbf{r},t)$ corresponds to the total 
appearance rate across the entire Galaxy at time $t$.

To model the scenario in which contacts with technoemissions have occurred over the past $\tau=65$ yr,
we start by considering an isotropic emitter that, at time $t$, generated a technoemission lasting for a duration $L$, propagating uniformly in all 
directions at the speed of light $c$. At the present time, the region covered by the radiation forms a spherical shell centered at $\mathbf{r}$, with 
outer and inner radii given by $ct$ and $\max(ct-cL,0)$,\footnote{For $t< L$, the inner radius is zero, meaning that the radiation fills a spherical region of radius $ct$}, respectively (Fig. \ref{fig1}a). 
As illustrated in Fig. \ref{fig1}b, $\tau$ years ago this same spherical shell had outer and inner radii smaller by $c\tau$. If, during this interval, the distance to the emitter lay between the inner radius from $\tau$ years ago, $\max(ct-cL-c\tau,0)$, and the current outer radius
$ct$, then the technoemission would have intersected Earth at some point within the past $\tau$ years (Fig. \ref{fig1}c).
This translates into the following condition on the appearance time:
\begin{equation}
\label{time}
d/c\leq t \leq d/c+L+\tau,
\end{equation}
where $d=\vert \mathbf{r}-\mathbf{r}_o\vert$ denotes the distance between the emitter and Earth, located at $\mathbf{r}_o$.

Integrating $\gamma(\mathbf{r},t)$ over $t$ and $\mathbf{r}$ under the constraint imposed by Equation \eqref{time} yields the mean 
number of contacts with technoemissions of duration $L$ that have occurred within the past $\tau$ years.
Assuming that the technoemission lifetimes follow a probability distribution function (PDF) $\rho_L(L)$,
and focusing on emitters located within a distance $R$ from Earth, the corresponding expected number $\eta(\tau,R)$ of technoemissions
intersecting Earth is given by:
\begin{equation}
\label{naverage1}
\eta(\tau,R)=\int\!dL \rho_L(L)\int\!d\mathbf{r}\int_{d/c}^{d/c+L+\tau}\! dt\,\gamma(\mathbf{r},t)\theta(R-d)
\end{equation}
where $\theta(x)$ is the Heavyside step function.

The isotropic technoemission model described above applies to Dysonian megastructures \citep{Dyson1960}, which
are expected to radiate in the infrared, and to other omnidirectional radio or optical signals.
Spectral technosignatures from the atmospheres of exoplanets can be considered 
another example of omnidirectional technoemissions \citep{Lin2014,Kopparapu2021,Haqq-Misra2022a}, as their transmitted or reflected 
starlight is not directed toward specific targets. 
However, we do not count them among the potentially missed signals
throughout SETI history, since searches for them have just become barely possible \citep{Seager2025}. Likewise,
we do not consider technoemissions produced by modifications of exoplanetary surfaces through intensive artificial 
illumination \citep{Beatty2022} or selectively engineered absorption/reflection of starlight \citep{Lingam2017a,Berdyugina2019,Jaiswal2023}, 
as no dedicated searches have yet been conducted \citep{Haqq-Misra2022b}. 

The present omnidirectional model for technoemissions can be
generalized to include directional signals, such as those produced by unidirectional beacons targeting other planetary 
systems, interstellar propulsion \citep{Benford2016,Lingam2017b}, or communication between spacecrafts \citep{Bracewell1960}. 
In this case, contact would occur only with those beacons whose beamwidth intersects Earth.
Here, we make the simplifying assumption that the
orientation and beamwidth of such directional signals are independent of the emitters' time of appearance, position, 
and longevity. Under this assumption, their contribution can be encoded by an anisotropic prefactor $\chi \leq 1$ applied to the right-hand side of Equation \eqref{naverage1}, reflecting the lower probability for a directional technoemission to intersect Earth's line of sight compared
to an isotropic one with the same occurrence rate and longevity \citep{Grimaldi2017,Grimaldi2023}.

In writing the probability distribution of contacts, we do not impose a strict upper limit on the population of emitters in the Galaxy, 
requiring only that the number of contacts remains much smaller than the estimated $\sim 10^{10}$ potentially habitable planets 
in the Milky Way \citep{Bryson2021}\footnote{By ``habitable planets" we mean rocky worlds orbiting within the habitable 
zones of FGK stars}. In this way, owing to the statistical 
independence of the emitters, the probability that exactly $n$ technoemissions have intersected Earth over the past
$\tau$ years from within a distance $R$ follows a Poisson distribution \citep{Grimaldi2023,Grimaldi2018}:
\begin{equation}
\label{poisson1}
P(n;\tau,R)=\frac{[\eta(\tau,R)]^n}{n!}e^{-\eta(\tau,R)}.
\end{equation}

\subsection{Stationary regime}
Proceeding further requires modeling the rate per unit volume, $\gamma(\mathbf{r},t)$, for which we lack prior knowledge, except from the requirements that emitter positions must lie within the Milky Way and that their appearance times cannot precede the birth of the Galaxy itself ($\sim 13.8$ Gyr ago). Here, we focus primarily on the instructive case of a stationary regime, in which $\gamma(\mathbf{r},t)$ is independent 
of the appearance time $t$. This assumption is justified by noting that, in Equation \eqref{time}, the interval during which technoemissions 
intersect Earth is $ L+ \tau$: if $\rho_L(L)$ is such that, within this interval, the time dependence of $\gamma(\mathbf{r},t)$ is sufficiently weak or negligible, then $\gamma(\mathbf{r},t)$ can be approximated as $\varGamma\rho_E(\mathbf{r})$, where $\varGamma$ is the appearance rate
of emissions across the entire Galaxy and $\rho_E(\mathbf{r})$ denotes the PDF of the emitter positions, normalized to unity. 
Within this stationary regime, the integration over $t$ in Equation \eqref{naverage1} yields $\eta(\tau,R)=\pi_R\chi\varGamma (\bar{L}+\tau)$, 
so Equation \eqref{poisson1} reduces to:
\begin{equation}
\label{poisson2}
P(n;\varGamma,\tau,\pi_R)=\frac{[\pi_R\chi\varGamma (\bar{L}+\tau)]^n}{n!}e^{-\pi_R\chi\varGamma (\bar{L}+\tau)}
\end{equation}
where $\bar{L}=\int\!dL\,\rho_L(L)L$ is the average longevity of the technoemissions and
\begin{equation}
\label{pio}
\pi_R=\int\!d\mathbf{r}\rho_E(\mathbf{r})\theta(R-\vert\mathbf{r}-\mathbf{r}_o\vert)
\end{equation} 
represents the probability of an emitter being within a distance $R$ from Earth \citep{Grimaldi2017}. Regardless of the specific 
distribution of emitters, $\pi_R$ increases monotonically with $R$, from $\pi_R=0$ at $R=0$ and reaching unity as $R$ becomes larger than the characteristic size of the Milky Way.
 
\subsection{Bayesian analysis}
Our objective is to infer the emission rate, $\varGamma$, from the hypothetical scenario, denoted $\mathcal{U}_{n,\tau}$, in which $n$ undetected 
contacts have occurred over the past $\tau$ years. To this end, we adopt a Bayesian inference approach, using Equation \eqref{poisson2} as the basis for constructing the likelihood function of $\mathcal{U}_{n,\tau}$ given $\varGamma$. In so doing, we note that the working hypothesis -- 
namely, that these $n$ contacts have gone undetected -- implies that the 
associated technoemissions were such as to elude our search efforts.
This could be due to various factors, such as their wavelength, position in the sky, or other characteristics, but
in particular because their signals may have been too faint or emitted from distances beyond the sensitivity limits of our telescopes.
In such a scenario, the undetected technoemissions could have originated anywhere within the Galaxy. The corresponding
likelihood function, $P(\mathcal{U}_{n,\tau}\vert \varGamma)$, is therefore provided by Equation \eqref{poisson2} with $\pi_R=1$:
\begin{equation}
\label{likelhood1}
P(\mathcal{U}_{n,\tau}\vert \varGamma)=P(n;\varGamma,\tau,1). 
\end{equation}

Bayes' theorem allows us to derive the posterior PDF of $\varGamma$ under the assumption that the scenario $\mathcal{U}_{n,\tau}$ has occurred: 
\begin{equation}
\label{bayes1}
p(\varGamma\vert \mathcal{U}_{n,\tau})=\dfrac{P(\mathcal{U}_{n,\tau}\vert \varGamma)p(\varGamma)}
{\int\!d\varGamma\,P(\mathcal{U}_{n,\tau}\vert \varGamma)p(\varGamma)},
\end{equation}
where $p(\varGamma)$ is the prior PDF encoding our assumptions about the possible values of $\varGamma$ before taking
the $n$ contacts into account. In what follows, we assume complete prior ignorance and adopt a log-uniform 
prior, $p(\varGamma)\propto 1/\varGamma$, which
assigns equal weight to all orders of magnitude of $\varGamma$. 
Under this assumption, the posterior PDF takes the form of a Gamma distribution for $\varGamma$: 
\begin{equation}
\label{bayes2}
p(\varGamma\vert \mathcal{U}_{n,\tau})=\chi(\bar{L}+\tau)\frac{[\chi\varGamma(\bar{L}+\tau)]^{n-1}}{(n-1)!}e^{-\chi\varGamma(\bar{L}+\tau)}.
\end{equation}
The corresponding posterior expectation value of the technoemission rate is then (Appendix \ref{AppA}):
\begin{equation}
\label{averageG}
\langle \varGamma\rangle =\chi^{-1}\frac{n}{\bar{L}+\tau},
\end{equation}
which increases with the number $n$ of past contacts and decreases as
the average longevity $\bar{L}$ of technosignatures increases. This behavior has a clear physical interpretation, stemming directly from
the contact condition in Equation \eqref{time}: the appearance time interval during which a technoemission of average duration $\bar{L}$ can
intersect Earth is $\bar{L}+\tau$. Consequently, $n$ contacts within the same interval must occur at a rate scaling as $n/(\bar{L}+\tau)$. The prefactor $\chi^{-1}\ge 1$ in Equation \eqref{averageG} compensates for the reduced intersection probability of anisotropic technoemissions.

\begin{figure*}[t]
	\begin{center}
		\includegraphics[width=0.95\textwidth,clip=true]{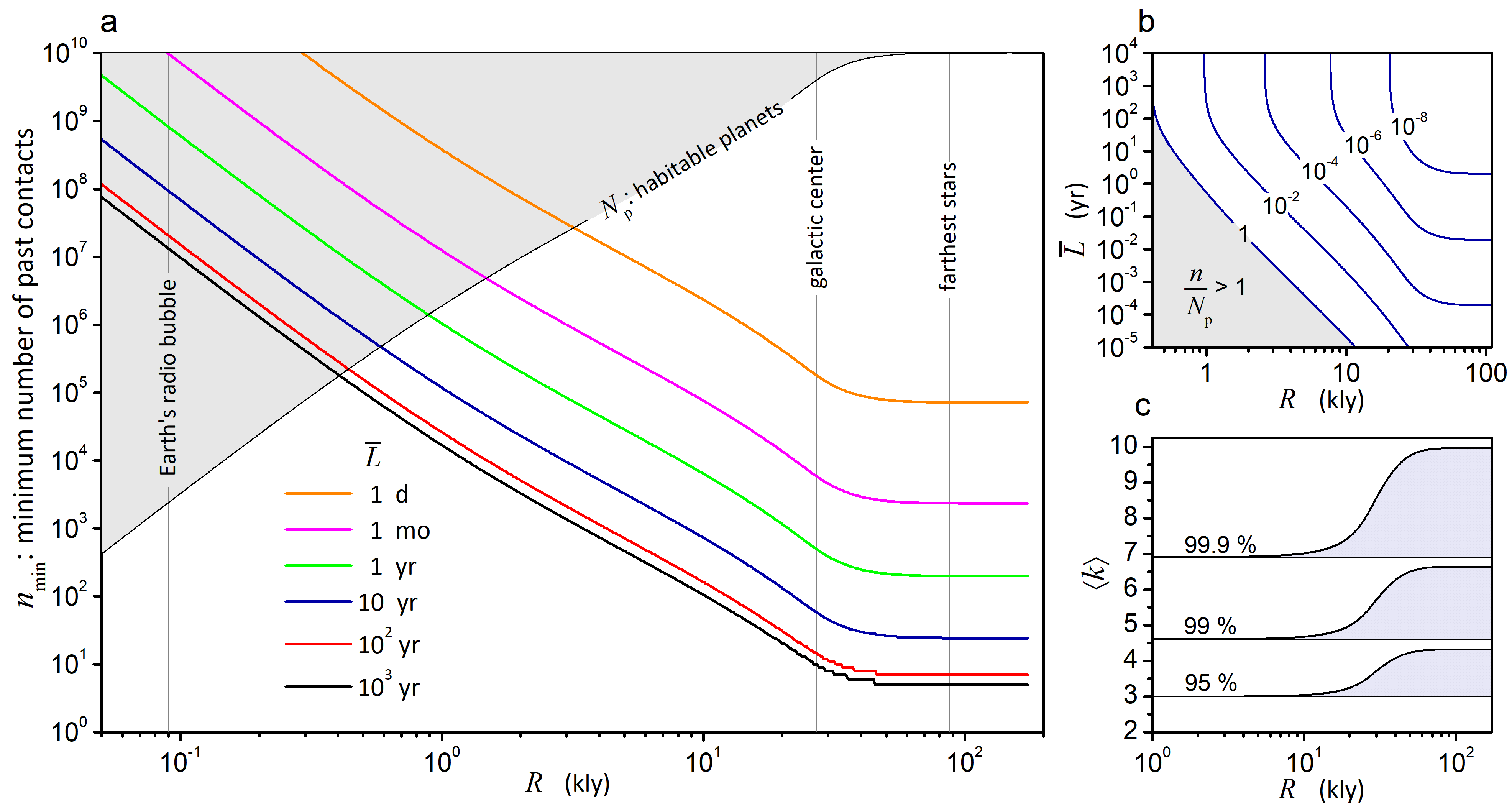}
		\caption{Results assuming undetected contacts with technoemissions over the past $65$ years. a: Number $n$ of undetected 
		past contacts required to achieve a present-day detectability of at least $95$ \%, as a function of the 
		search radius $R$. The thin black line shows the number of habitable planets $N_p$ as a function of R calculated from 
		$N_p=10^{10}\int\!d\mathbf{r}\rho_\textrm{disk}(\mathbf{r})\theta(R-d)$, where $d$ is the distance from Earth. The grey region
		marks where $n>N_p$. b: Average emitter longevity
		versus $R$ for selected $n/N_p$ values. c: Upper and lower bounds on the expected number of present-day 
		detections, $\langle k\rangle$, computed from Equation \eqref{bounds} for detectabilities $\mathcal{P}=95$\%, $99$\%, and $99.9$\% . 
		For each $\mathcal{P}$, all possible $\langle k\rangle$ values lie within the corresponding colored areas.
		}\label{fig2}
	\end{center}
\end{figure*}

\section{Results and discussion}
\subsection{Present-day detectability of technoemissions}
Having established the impact of past undetected contacts on the rate of technoemissions, we next explore how such past contacts 
influence technoemission detectability today. 
By ``detectability",  we mean the capability of present-day and near-future telescope technologies to detect a technoemission. 
This capability has considerably advanced in recent years, particularly in terms 
of the range of searched wavelengths, sensitivity thresholds, and survey speeds \citep{Tremblay2024, Manunza2025,Mason2025,Maire2022,Luan2023,Tremblay2025}. 
To appreciate the importance of recent efforts, 
it suffices to note that since its launch in 2016, Breakthrough Listen has explored a parameter search space largely exceeding that of all  
previous searches combined over the preceding fifty years \citep{Garrett2023}. 
Looking ahead, upcoming facilities -- such as the Square Kilometre 
Array (SKA) and the Next Generation Very Large Array (ngVLA) -- 
promise to significantly enhance the observational prospects for technosignature science \citep{Siemion2015,Ng2022}.

In the following, we define a technoemission intersecting Earth as detectable if its emitter lies within a distance $R$ from our planet.
This criterion conveniently encapsulates both the intrinsic properties of the emission and the instrumental detection capabilities within 
a single parameter. 
A radio or microwave signal with an effective isotropic radiated power $W$ is potentially detectable up to a distance
$R =\sqrt{W/4\pi S_\textrm{min}}$,
where $S_\textrm{min}$ denotes the minimum detectable flux, determined by factors such as bandwidth, integration time, and other instrumental parameters \citep{Enriquez2017}. For optical or infrared technoemissions, the maximum detectable 
distance $R$ is likewise determined by the intrinsic properties of the emission as well as by instrumental specifications \citep{Howard2004}.

To account for the diversity of such specifications and emission characteristics, $R$ is treated here as a 
free parameter. 
Accordingly, Equation \eqref{poisson2} with $\pi_R\le 1$ and $\tau=0$ yields the probability that exactly $k=0,\,1,\,2,\,\ldots$ technoemissions,
generated at a constant rate $\varGamma$, are detectable at the present time (event $\mathcal{D}_k$) \citep{Grimaldi2018}:
\begin{equation}
\label{likelihood2}
P(\mathcal{D}_k\vert \varGamma)=P(k; \varGamma, 0,\pi_R).
\end{equation}
Marginalizing $P(\mathcal{D}_k\vert \varGamma)$
with respect to the posterior PDF of $\varGamma$  yields the posterior distribution of $k$ detectable 
contacts, conditional on the occurrence of $n\geq 1$ undetected contacts over the past $\tau$ years:
\begin{align}
\label{probkn}
P(\mathcal{D}_k\vert \mathcal{U}_{n,\tau})&=\int_0^\infty\!d\varGamma P(\mathcal{D}_k\vert \varGamma)p(\varGamma\vert \mathcal{U}_{n,\tau})\nonumber \\
&=\frac{(\pi_R\bar{L})^k(\bar{L}+\tau)^n}{[\bar{L}(1+\pi_R)+\tau]^{k+n}}\frac{(n+k-1)!}{k!(n-1)!}.
\end{align}
Notably, $P(\mathcal{D}_k\vert \mathcal{U}_{n,\tau})$ is independent of the anisotropy factor $\chi$ (Appendix \ref{AppB}). 
This arises because the detection probability of a directional signal is reduced by a factor $\chi$, exactly compensating for the
 $\chi^{-1}$ enhancement of its occurrence rate.
Consequently, all results derived from \eqref{probkn} apply equally to both isotropic and directional 
technoemissions, as well as to a mixture of them\footnote{This may not hold true when considering mixtures of 
isotropic and anisotropic signals having differing longevities.}.

We define the present-day detectability as the probability $\mathcal{P}$ of at least one detectable contact occurring today:
\begin{equation}
\label{prob1}
\mathcal{P}=1-P(\mathcal{D}_{k=0}\vert \mathcal{U}_{n,\tau})=1-\left(\frac{\bar{L}+\tau}{\bar{L}(1+\pi_R)+\tau}\right)^n,
\end{equation}
For any given value of the search radius $R$ and the average longevity $\bar{L}$, $\mathcal{P}$ is a strictly increasing function of $n$.
While this aligns with our intuition -- the more contacts there have been in the past, the greater the probability of detection today --
it also raises a more intriguing and relevant question for technosignature science: How many contacts must have gone undetected over 
the past $\tau=65$ years for there to be a high probability of detection today?

\begin{figure*}[t]
	\begin{center}
		\includegraphics[width=0.9\textwidth,clip=true]{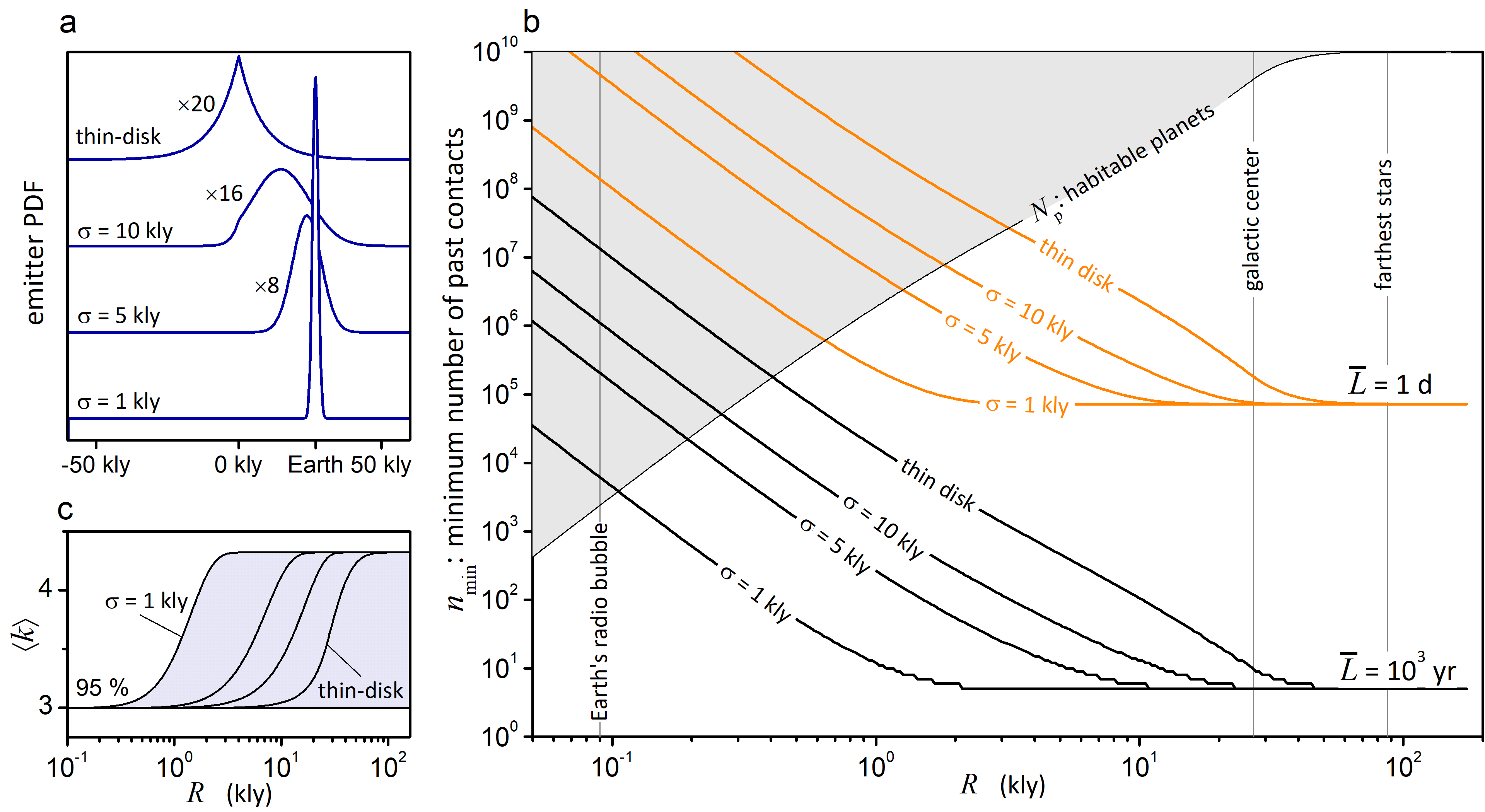}
		\caption{Effects of emitter clustering around Earth on past undetected contacts.  a: Profile of the
		emitter PDF as a function the radial distance from the galactic center for different values of the standard 
		deviation $\sigma$. b: Number $n$ of undetected past contacts required to achieve a present-day detectability $\mathcal{P}$ 
		of at least $95$ \%, as a function of the search radius $R$. c: Upper and lower bounds on $\langle k\rangle$ computed
		from Equation \eqref{bounds} for $\mathcal{P}= 95$ \%}\label{fig3}
	\end{center}
\end{figure*}

Figure \ref{fig2} provides an initial answer by showing the minimum number $n_\textrm{min}$ of past undetected 
contacts required to ensure at least a $95$ \% probability that technosignals are detectable within a distance $R$ from Earth. 
$n_\textrm{min}$ is readily obtained by inverting Equation \eqref{prob1}:
\begin{equation}
\label{probkn3}
n_\textrm{min}=\left\lceil\dfrac{\ln(1-\mathcal{P})^{-1}}{\ln\left(\dfrac{\bar{L}(1+\pi_R)+\tau}{\bar{L}+\tau}\right)}\right\rceil,
\end{equation}
where $\lceil \cdots\rceil$ denotes the ceiling function, showing that $n_\textrm{min}$ reaches its minimum value
when $\pi_R\rightarrow 1$, whereas it becomes unbounded in the limit $\pi_R\rightarrow 0$ (i.e., $R\rightarrow 0$).
The calculations shown in Fig. \ref{fig2} assume that 
the distribution of emitters, $\rho_E(\mathbf{r})$, follows that of stars in the galactic thin disk, denoted $\rho_\textrm{disk}(\mathbf{r})$, 
which is modeled by a double exponential with a radial scale length of $2.5$ kpc ($8.15$ kly) and a vertical scale length of 
$0.16$ kpc ($0.52$ kly) \citep{Licquia2016,McMillan2017}.

In Figure \ref{fig2}a, $n_\textrm{min}$ is lowest when the search radius covers 
the entire galactic disk ($R\gtrsim 60$ kly), ranging from $n_\textrm{min}=5$ for $\bar{L}\ge 10^3$ yr to 
as many as $n_\textrm{min} \simeq10^5$ for 
technoemissions lasting only one day on average -- and even higher for shorter durations. 
These values remain far smaller than the estimated $\sim 10^{10}$ habitable planets around FGK stars, which we take as a proxy for 
the maximally conceivable -- albeit highly implausible -- number of potential emitters. However, as the search radius
decreases, the number of undetected contacts rises steadily to maintain $\mathcal{P} \ge 95$ \%, eventually surpassing even the number
$N_p=10^{10}\int\!d\mathbf{r}\rho_\textrm{disk}(\mathbf{r})\theta(R-d)$ of habitable planets within the same range.

This provides a first important insight into our chances of detecting technosignatures. 
Even under the overly optimistic assumption of one 
emitter per $100$ habitable planets, the results shown in Figs. \ref{fig2}a and \ref{fig2}b imply that detectable (i.e., $\mathcal{P}\ge 95$\%)
long-lived emitters must lie at distances beyond $\sim 1,000$ ly from Earth, or beyond $\sim 10,000$ ly in the case of short-lived ones. 
These thresholds correspond to $\sim 10^4$ long-lived undetected contacts over the past $65$ years 
(i.e., roughly one every two days) or $\sim 10^6$ short-lived ones (about 40 per day). More conservative -- and arguably more 
plausible -- rates of past contacts would push these minimal distances even farther.
For example, less than one undetected contact per year ($n_\textrm{min}\le 65$)  would be achieved
only for $R\gtrsim 12,000$ ly in the case of long-lived technoemissions. 

A less stringent detectability requirement does not significantly alter these conclusions. For example, at fixed
$R$, reducing the detectability threshold in Equation \eqref{probkn3}  from $\mathcal{P}\ge 95$\% to $\mathcal{P}\ge 50$\% decreases $n_\textrm{min}$ by only a factor of $\sim 4$.

\subsection{Average number of present-day detectable technoemissions}
A second key result pertains the expected number of present-day detectable technoemissions, given $n\ge 1$ past contacts. This is obtained by 
summing $k P(\mathcal{D}_k\vert \mathcal{U}_{n,\tau})$ over all $k\ge 0$, yielding (Appendix \ref{AppC}):
\begin{equation}
\label{kbar} \langle k\rangle = \frac{\pi_R\bar{L}}{\bar{L}+\tau}n,
\end{equation}
implying that $\langle k\rangle$ growns indefinitely with the number of past contacts. However,
when combined with Equation \eqref{prob1}, it can be shown that the expected number of detectable technoemissions is bounded 
as (Appendix \ref{AppC}):
\begin{equation}
\label{bounds}
\ln(1-\mathcal{P})^{-1}\le\langle k\rangle \le \frac{\pi_R\ln(1-\mathcal{P})^{-1}}{\ln(1+\pi_R)} ,
\end{equation}
showing that $\langle k\rangle$ actually increases only logarithmically with the detection probability $\mathcal{P}$.
Even in the case of almost certain detectability (e.g., $\mathcal{P}=99.9$ \%), $\langle k\rangle$ remains below $10$, 
regardless of the search radius (Fig. \ref{fig2}c). 
Thus, while increasing $R$ helps reduce the number of undetected past contacts to more conservative values, it also greatly 
expands the search volume -- potentially to galactic scales -- within which, however, only a 
few detectable technoemissions might be expected among an astronomically large number of stars.

This trade-off points to an optimal search radius of at least a few thousand light-years, a range that is, in principle, accessible 
to the FAST telescope \citep{Luan2023} or the eventual completion of the SKA \citep{Siemion2015} and the ngVLA \citep{Ng2022}, 
provided the technoemissions are sufficiently powerful (at least an EIRP equivalent to that of the Arecibo telescope) and long-lived.

\begin{figure*}[t]
	\begin{center}
		\includegraphics[width=0.9\textwidth,clip=true]{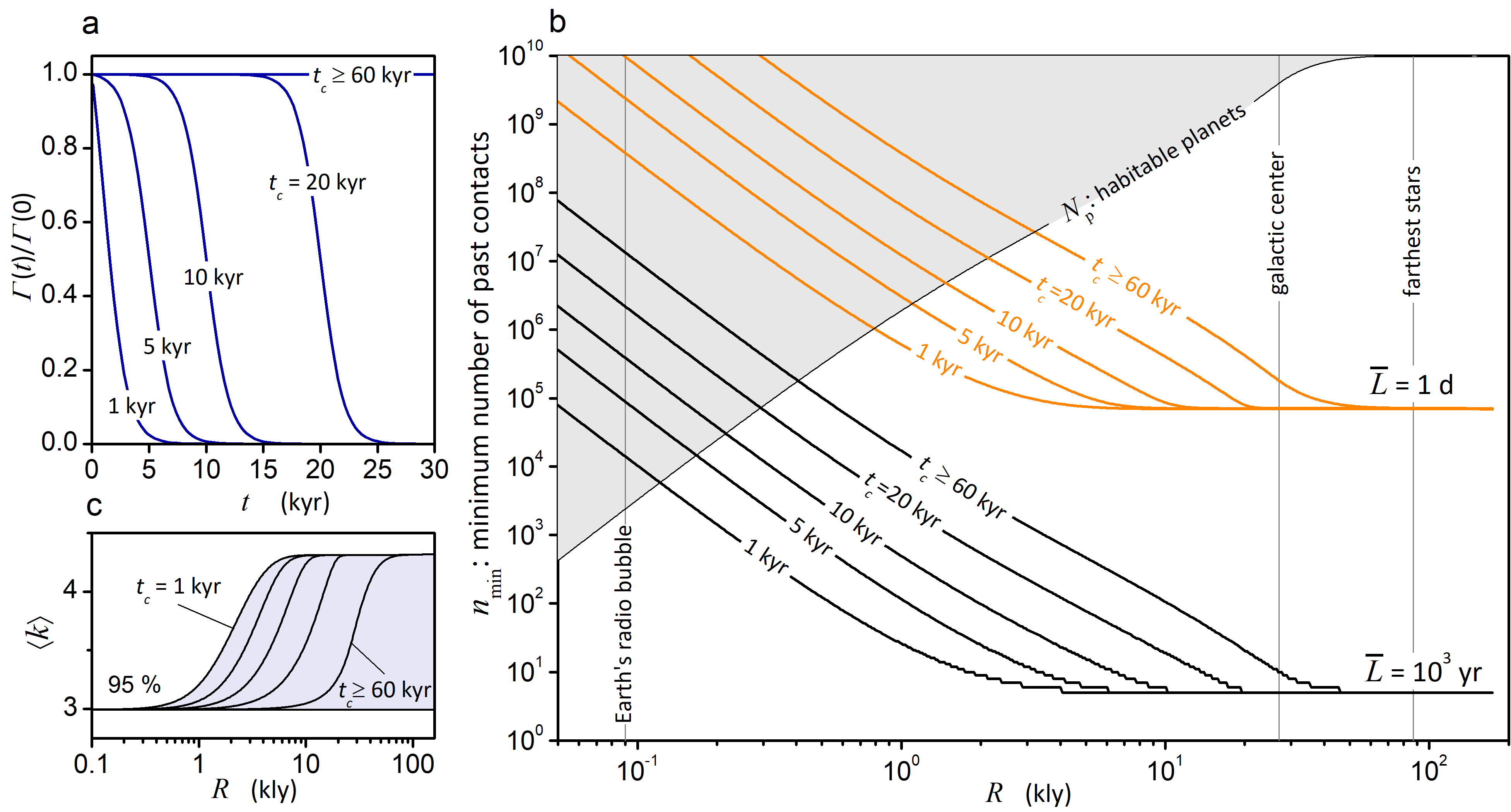}
		\caption{Effects of a temporal transition toward a high emitter birthrate on the number of undetected past contacts. 
		a: Dependence of emitter appearance time for different transition times $t_c$. The growth rate is fixed at $1$ kyr$^{-1}$. 
		b: Corresponding minimum number $n_\textrm{min}$ of undetected past contacts required to 
		achieve a present-day detectability $\mathcal{P}$ of at least $95$ \%. c: Upper and lower bounds on $\langle k\rangle$ computed
		from Equation \eqref{bounds} for $\mathcal{P}= 95$ \%.}\label{fig4}
	\end{center}
\end{figure*}

\subsection{Insensitivity to the prior}
The requirement of past contacts, encoded in the likelihood function, makes the above results 
largely insensitive to the choice of the prior, even for informative ones. For instance, adopting
a prior PDF for $\varGamma$ of the form 
$p(\varGamma)\propto \varGamma^{\alpha-1}$ -- ranging from optimistic ($\alpha \simeq 1$) to pessimistic ($\alpha\simeq -1$)
 beliefs about the possible values of $\varGamma$ \citep{Spiegel2012,Grimaldi2023} -- would 
 only replace $n$ by $n+\alpha$ in the formulas for $\mathcal{P}$ and $\langle k\rangle$, leaving the results in Fig. \ref{fig2} 
 essentially unchanged for $n$ sufficiently large.  Thus, the conclusions drawn above remain 
 valid even under substantially different prior assumptions.   
 
\subsection{Effects of spatio-temporal dependences of emission rates}
A potential limitation of the above analysis stems from the assumption of a stationary rate of technoemissions from emitters uniformly distributed
across the thin disk of the Milky Way. 
Depending on the evolutionary history of technological species in our Galaxy, the spatio-temporal distribution of emitters may deviate significantly from this idealized scenario \citep{Cirkovic2004,Balbi2021}, leading to a broad range of possible configurations that could affect the conclusions drawn in the previous section.

Here, we focus on cases that allow for high present-day detectability at distances shorter than several thousand light-years,
without requiring an implausibly large number of past contacts. Our main objective is to assess to what extent these cases represent 
a viable alternative to the results shown in Fig. \ref{fig2}.

We begin by examining a significant deviation from a uniform emitter distribution across the thin disk, which could 
result from regions of the Galaxy more favorable to the development of advanced civilizations or from higher local prevalence
of space probes.
Specifically, we consider emitters to be more or less localized around Earth, following a PDF of the form 
$\rho_E(\mathbf{r})\propto \rho_\textrm{disk}(\mathbf{r})e^{-d^2/2\sigma^2}$, where, as before, $\rho_\textrm{disk}(\mathbf{r})$ 
is the spatial PDF of stars in the thin disk.
The rationale for this choice is that for a high local concentration of emitters (i.e., small $\sigma$), the number of past contacts
required to achieve a given detectability $\mathcal{P}$
is expected to level off well before $R$ reaches the scale of the Galactic disk, since the emitter distribution decreases rapidly
beyond $R\sim \sigma$. 

This trend is confirmed by the results shown in Fig. \ref{fig3}b: $n_\textrm{min}$ decreases systematically as emitters become more 
localized around Earth. 
However, achieving both high detectability ($\mathcal{P}\ge 95$ \%) and more plausible numbers of past contacts at relatively short
distances requires emitters to be tightly clustered around Earth. For instance, fewer than one
long-lived past contact per year ($n_\textrm{min}\lesssim 65$) is attainable for $R\gtrsim 500$ ly if almost all emitters are 
long-lived and located within $1,000$ light years of Earth (Fig. \ref{fig3}a), corresponding to merely $\sim 0.02$\% 
of the thin-disk volume. 
Even tighter emitters localizations would be required to attain similar frequencies of past contacts from short-lived emitters.
Notably, the bounds on the present-day number of detectable technoemissions
remain relatively unaffected by the localization parameter $\sigma$ (Fig. \ref{fig3}c).

Another hypothetical scenario that could yield realistic values of $n$ on length scales well below that of the Galaxy is one in which the 
emitter birthrate has increased in the relatively recent past \citep{Cirkovic2004}. 
Such an increase could result from large-scale colonization 
events \citep{Lingam2016,Carroll2019} or from the abandonment of a non-interference directive  -- such as the one envisioned by the zoo hypothesis \citep{Ball1973, Crawford2024} -- in which advanced 
alien civilizations deliberately avoid contact to preserve the natural development of less advanced societies. 
A secular increase of the emitter birthrate may also result from astrobiological phase transition 
scenarios \citep{Cirkovic2008}, in which
the gradual decline of global catastrophic events such as gamma-ray bursts \citep{Annis1999} may have only recently
allowed the emergence of technological species in the Milky Way. 

Although the aforementioned scenarios describe some specific mechanisms for a temporal increase of the emitter birthrates,
the possible causes and the details of such a transition are of secondary importance for our analysis. Here, we consider
a simplified model in which the Galaxy is virtually devoid of emitters 
for $t>t_c$ and becomes populated for $t<t_c$, where $t_c$ denotes the timescale of the transition. 
To this end, we adopt a separable form for the appearance 
rate per unit volume, $\gamma(\mathbf{r},t)=\varGamma(t)\rho_\textrm{disk}(\mathbf{r})$, where
$\varGamma(t)\propto 1/[e^{\beta(t-t_c)}+1]$ is normalized such that $\varGamma(t=0)=\varGamma$.
For illustration, we set the growth rate to $\beta=1$ kyr$^{-1}$ (Fig. \ref{fig4}a). 

Figure \ref{fig4}b shows that the
number of past contacts from emitters located beyond $\sim ct_c$ quickly reaches its lower bound as a function of $R$, in 
analogy with the spatial localization case considered above. However, achieving a significant reduction of $n_\textrm{min}$, 
while maintaining a high detectability today, requires a very recent increase in the emitter population. 
For instance, to achieve $n_\textrm{min}\lesssim 65$ for $R\gtrsim 600$ ly in the case of long-lived 
technoemissions, the transition must have occurred only $\sim 1$ kyr ago -- that is, during historical times for humanity -- with no significant change in $\langle k\rangle$ (Fig. \ref{fig4}c). Growth rates slower than $\beta\sim 0.1$ kyr$^{-1}$, as would be expected for 
large-scale colonization or astrophysical phase transitions, would effectively 
recover the stationary scenario of Fig. \ref{fig2}.

The key result from Figs. \ref{fig3} and \ref{fig4} is that avoiding untenably large numbers of past contacts at small search 
radii requires invoking no less implausible scenarios, such as emitters confined to our immediate vicinity or a very recent and sudden
increase in their population. These findings reinforce our conclusion that assuming undetected contacts throughout SETI's history does not 
necessarily imply a high detection probability today ($\mathcal{P}\ge 95$\%), unless emitters are long-lived and located several 
thousand light-years from Earth. 
Even then, the expected number of detectable technoemissions increases only 
logarithmically with $\mathcal{P}$, suggesting that, even with near-certain detectability and precise knowledge of what 
to search for across the vast parameter search space, at most a few would be expected in the entire Milky Way.

Before concluding, a few remarks are in order regarding the general applicability of our model to scenarios other
than the distributions of emitters in our Galaxy considered here. In principle, our formalism could be adapted to account for 
intergalactic technoemissions. In such a case, however, the emitter distribution $\rho_E(\mathbf{r})$ would need to be appropriately 
modeled, and the resulting estimates of the number of past undetected contacts would have to be compared to plausibility 
parameters different from the number $N_p$ of habitable exoplanet in our Galaxy. Likewise, in considering the possibility of emitters in our immediate neighborhood -- particularly technoemissions 
from artifacts in the Solar System or within a few light-years from Earth -- $N_p$ would cease to be a meaningful parameter for 
assessing the plausibility of such a scenario, and alternative benchmarks would need to be considered.

\begin{acknowledgments}
CG is grateful to the referee for constructive suggestions that significantly improved the manuscript, and for helpful discussions 
with Amedeo Balbi, Paolo De Los Rios, and Huckleberry David Gaston Thums.
\end{acknowledgments}

\appendix
\section{Conditional expectation of the emission rate} 
\label{AppA}
The expected value of $\varGamma$, conditional on $n\ge 1$ past contacts, is obtained from the posterior PDF in \eqref{bayes2}:
\begin{equation}
\label{avegamma1}
\langle \varGamma\rangle = \int_0^\infty \! d\varGamma\, \varGamma p(\varGamma\vert \mathcal{U}_{n,\tau}) 
=\int_0^\infty \! d\varGamma\,\frac{[\chi\varGamma(\bar{L}+\tau)]^n}{(n-1)!}e^{-\chi\varGamma(\bar{L}+\tau)}.
\end{equation}
After changing the integration variable to $x=\chi\varGamma (\bar{L}+\tau)$ and using $\int_0^\infty\! dx x^n e^{-x}=n!$,
\eqref{avegamma1} reduces to:
\begin{equation}
\label{avegamma2}
\langle \varGamma\rangle=\chi^{-1}\frac{n}{\bar{L}+\tau}.
\end{equation}

\section{independence of the present-day detectability on technoemission anistropy}
\label{AppB}
The probability of $k$ detectable contacts, conditional on the occurrence of $n\geq 1$ undetected contacts over the past $\tau$ years, 
is given by the following predictive posterior distribution:
\begin{equation}
\label{probkn0}
P(\mathcal{D}_k\vert \mathcal{U}_{n,\tau})=\int_0^\infty\!d\varGamma P(\mathcal{D}_k\vert \varGamma)p(\varGamma\vert \mathcal{U}_{n,\tau})
=\dfrac{1}{k!(n-1)!}\int_0^\infty\!d\varGamma (\chi\pi_R\varGamma \bar{L})^k
\chi(\bar{L}+\tau)[\chi\varGamma (\bar{L}+\tau)]^{n-1}e^{-\chi\varGamma [\bar{L}(1+\pi_R)+\tau]}.
\end{equation}
Setting the integration variable to $x=\chi\varGamma$ makes the integral independent of the anisotropy factor $\chi$. 
This stems from a perfect compensation: the $\chi^{-1}$-amplification of the emission rate \eqref{avegamma2} is 
canceled by the factor $\chi$ in the detection probability. A further change of the integration variable then yields: 
\begin{equation}
\label{probkn1}
P(\mathcal{D}_k\vert \mathcal{U}_{n,\tau})=\frac{(\pi_R\bar{L})^k(\bar{L}+\tau)^n}{[\bar{L}(1+\pi_R)+\tau]^{k+n}}
\dfrac{\int_0^\infty\! dx\, x^{n+k-1}e^{-x}}{k!(n-1)!}
=\frac{(\pi_R\bar{L})^k(\bar{L}+\tau)^n}{[\bar{L}(1+\pi_R)+\tau]^{k+n}}
\frac{(n+k-1)!}{k!(n-1)!}.
\end{equation}

\section{Bounds on $\langle k\rangle$}
\label{AppC}
The average number of present-day detectable technoemissions, conditioned to $n$ past contacts, is obtained from:
\begin{equation}
\label{avek1}
\langle k\rangle=\sum_{k=0}^\infty k\,P(\mathcal{D}_k\vert \mathcal{U}_{n,\tau})=\sum_{k=0}^\infty k\! \int_0^\infty\!d\varGamma P(\mathcal{D}_k\vert \varGamma)p(\varGamma\vert \mathcal{U}_{n,\tau}).
\end{equation}
Moving the sum over $k$ inside the integral and using \eqref{avegamma1} and \eqref{avegamma2} yields:
\begin{equation}
\label{avek2}
\langle k\rangle =\int_0^\infty\!d\varGamma \left[\sum_{k=0}^\infty k P(\mathcal{D}_k\vert \varGamma)\right]p(\varGamma\vert \mathcal{U}_{n,\tau}) 
=\chi\pi_R\bar{L}\langle \varGamma\rangle=\frac{\pi_R\bar{L}}{\bar{L}+\tau}n.
\end{equation}
To find upper and lower bounds on $\langle k\rangle$, we use \eqref{prob1} to express $n$ in terms of the detection probability 
$\mathcal{P}$ and rewrite \eqref{avek2} as:
\begin{equation}
\label{avek3}
\langle k\rangle=\frac{\pi_R\bar{L}}{\bar{L}+\tau}\dfrac{\ln(1-\mathcal{P})^{-1}}{\ln\!\left(1+\dfrac{\pi_R\bar{L}}{\bar{L}+\tau}\right)}.
\end{equation}
For fixed $\mathcal{P}$, $\langle k\rangle$ is a continuous increasing function of $\bar{L}$, with lower and upper limits given respectively by:
\begin{equation}
\begin{array}{ccccc}
\lim_{\bar{L}=0}\langle k\rangle=\ln(1-\mathcal{P})^{-1}, & & \textrm{and} & &
\lim_{\bar{L}=\infty}\langle k\rangle=\dfrac{\pi_R\ln(1-\mathcal{P})^{-1}}{\ln(1+\pi_R)}.
\end{array}
\end{equation}

\bibliography{manuscript-2}{}
\bibliographystyle{aasjournalv7}

\end{document}